\documentclass[preprint]{aastex631}

\received{}
\revised{}
\accepted{}
\submitjournal{ApJ}

\shorttitle{Electrostatic fluctuations in the magnetosheath}
\shortauthors{Perri et al.}
\graphicspath{{./}{figures/}}

\begin{document}

\title{Nature of electrostatic fluctuations in the terrestrial magnetosheath}

\correspondingauthor{Silvia Perri}
\email{silvia.perri@fis.unical.it}

\author[0000-0002-8399-3268]{Silvia Perri}
\affiliation{Dipartimento di Fisica, Universit\'a della Calabria\\
Ponte P. Bucci Cubo 31C \\
Rende, I-87036, Italy}

\author[0000-0003-1059-4853]{Denise Perrone}
\affiliation{ASI -- Agenzia Spaziale Italiana\\ Rome, Italy}

\author{Owen Roberts}
\affiliation{Space Research Institute\\ 
Graz, Austria}

\author{Adriana Settino}
\affiliation{Dipartimento di Fisica, Universit\'a della Calabria\\
Ponte P. Bucci Cubo 31C \\
Rende, I-87036, Italy}

\author{Emilya Yordanova}
\affiliation{Swedish Institute for Space Physics, \r{A}ngstr\"om Laboratory\\ 
Lägerhyddsvägen 1, 75121  Uppsala, Sweden}

\author[0000-0002-5981-7758]{Luca Sorriso-Valvo}
\affiliation{Swedish Institute for Space Physics, \r{A}ngstr\"om Laboratory\\
Lägerhyddsvägen 1, 75121  Uppsala, Sweden}
\affiliation{Istituto per la Scienza e Tecnologia dei Plasmi, Consiglio Nazionale delle Ricerche\\
via Amendola 122/D, 70126, Bari, Italy}

\author{Pierluigi Veltri}
\affiliation{Dipartimento di Fisica, Universit\'a della Calabria\\
Ponte P. Bucci Cubo 31C \\
Rende, I-87036, Italy}

\author{Francesco Valentini}
\affiliation{Dipartimento di Fisica, Universit\'a della Calabria\\
Ponte P. Bucci Cubo 31C \\
Rende, I-87036, Italy}



\begin{abstract}
The high cadence plasma, electric, and magnetic field measurements by the  Magnetospheric MultiScale spacecraft allow us to explore the near-Earth space plasma with an unprecedented time and spatial resolution, resolving electron-scale structures that naturally emerge from plasma complex dynamics. The formation of small-scale turbulent features is often associated to structured, non-Maxwellian particle velocity distribution functions that are not at thermodynamic equilibrium. Using measurements in the terrestrial magnetosheath, this study focuses on regions presenting bumps in the power spectral density of the parallel electric field at sub-proton scales. Correspondingly, it is found that the ion velocity distribution functions exhibit beam-like features at nearly the local ion thermal speed. Ion cyclotron waves in the ion-scale range are frequently observed at the same locations. 
These observations, supported by numerical simulations, are consistent with the generation of ion-bulk waves that propagate at the ion thermal speed. This represents a new branch of efficient energy transfer at small scales, which may be relevant to weakly collisional astrophysical plasmas.
\end{abstract}


\section{Introduction} \label{sec:intro}

The heating of weakly collisional interplanetary plasmas is nowadays a top priority subject in space physics. Due to the turbulent nature of such  systems \citep[see, e.g.][for the solar wind]{BrunoCarbone2013}, energy is efficiently carried from large to small scales, where significant departures of the plasma from the local thermodynamic equilibrium generates non-Maxwellian features in the particle velocity distribution function (VDF)~\citep[see e.g.][]{Marsch2006}. The identification of the fluctuation channels, along which this energy transfer occurs, represents a crucial step in the general problem of heating in nearly collisionless plasmas. 
Many theoretical and experimental efforts have been devoted to understand the physical mechanisms responsible for the local heating of the turbulent interplanetary plasma.
\citet{He15} have identified the occurrence of cyclotron and Landau resonances between kinetic fluctuations (consistent of quasi-parallel left-handed Alfv\'en-cyclotron waves and quasi-perpendicular right-handed kinetic
Alfv\'en waves) and the formation of the core-beam protons using high cadence WIND data. Analysis of \textit{in situ} data from the Helios spacecraft had shown the presence of significant electrostatic fluctuations in the high-frequency range (a few kHz) of the turbulent spectra ~\citep{GurnettAnderson1977,GurnettFrank1978,GurnettEA1979}. 
This electrostatic noise, when viewed on a time scale of several hours or more, is found during a large fraction (30-50\%) of the time~\citep{GurnettFrank1978} and it had been identified as of ion-acoustic nature.
More recent observational~\citep{MangeneyEA1999} and  numerical works~\citep{AranedaEA2008,MatteiniEA2010a,MatteiniEA2010b} proposed an analogous identification. However, the presence of this electrostatic activity even at small values of the electron to proton temperature ratio, $T_e/T_p$, cannot be explained within this scenario, since the ion-acoustic waves are heavily Landau-damped at low $T_e/T_p$~\citep{KrallTrivelpiece1986}.
Electrostatic fluctuations have routinely been observed in the magnetosheath, where several wave modes at sub-ion scales have been identified \citep{Zhu19}, and close to the Earth's bow shock. They are supposed to play an important role in electron thermalization close to the shock \citep[][]{Vasko18,Wang21}.

Recently, Hybrid Vlasov-Maxwell (HVM) numerical simulations of solar wind turbulence were used to analyze fluctuations longitudinal to an ambient magnetic field, at wavelengths shorter than the proton skin depth  $d_p=v_A/\Omega_{cp}$, where $v_A$ is the Alfv\'en speed and $\Omega_{cp}$ is the proton gyrofrequency~\citep{Valentini07,Valentini08,Valentini09,ValentiniEA2010,Valentini11b}.
These simulations showed that the resonant interaction of protons with left-handed polarized Alfv\'en-cyclotron fluctuations (at scales larger than $d_p$) generates diffusive plateaus (or small bumps) in the longitudinal proton VDF~\citep{KennelEngelmann1966}. For proton plasma beta (i.e. the ratio of the plasma pressure to the magnetic pressure) $\beta_p\sim 1$, such diffusive plateau is located near the proton thermal speed $v_{th,p}$~\citep{Valentini11b}. 
Due to the presence of the plateau, a nonlinear mode conversion of Alfv\'en-cyclotron waves takes place and a branch of electrostatic fluctuations can be excited \citep{Pegoraro20}.

Called ion-bulk (IBK) waves, these have phase speed close to $v_{th,p}$, acoustic-type dispersion and, at variance with the ion-acoustic waves, can survive against Landau damping even for small $T_e/T_p$~\citep{Valentini11b}, thanks to the presence of a region of flat proton VDF. Thus, the generation of such waves is triggered by the formation of ion beam/plateau of the VDF within its core (i.e., bulk).
Indeed, as shown numerically in \citet{ValentiniEA2010} the
electrostatic hybrid-Vlasov dielectric function admits two solutions for the wave phase speed when equilibrium distributions with velocity plateaus are considered: the ion-acoustic waves and the IBK waves. It has also been demonstrated that for $T_e/T_p\lesssim1$, only one branch exists, which propagates at almost the ion thermal speed \citep[][]{Valentini11}. 

The nature of these waves has extensively been investigated numerically~\citep{Valentini11,ValentiniEAPPCF2011}. They have also been detected at large wavenumbers in solar wind observations of the electric field signal from STEREO spacecraft~\citep{Valentini14b,VecchioEA2014}.

Thanks to the Magnetospheric MultiScale (MMS) constellation of spacecraft, capable to measure the near-Earth plasma at a cadence never reached by previous missions, it has been possible to explore in detail the kinetic processes coexisting with the nonlinear turbulent energy transfer from ion to electron scales.
While the four-point configuration has allowed to explore the structures forming in the plasma and to resolve reconnection sites down to the electron diffusion region \citep{Burch16,Eriksson16,Ergun16}, the high cadence plasma data have permitted to identify features emerging in both the ion and electron VDFs that show the effect of the interaction between particles and the electromagnetic turbulent fields \citep{Graham16,Graham17,Servidio17,Chasapis18,Chen19}. For example, it has been found that high coherence between electric field and electron bulk velocity fluctuations leads to the dissipation of kinetic and compressive Alfv\'enic turbulence in the magnetosheath \citep{He20}.

Recently, \citet{Sorriso2019} have pointed-out the effect of the energy transfer rate of the high-frequency turbulent cascade on the ion VDFs. When the turbulent transfer rate is locally high, non-Maxwellian features in the ion VDFs emerge. In particular, ion beams are forming in the presence of dominating Alfv\'enic fluctuations in the cascade and high electrostatic activity. 
Robust correlation between electric field fluctuations in the plasma frame and departure from Maxwellian ion VDFs has also been addressed in \citet{Perri2020}, suggesting that parallel and perpendicular electric field components play a role in the VDF deformation and the occurrence of non-linear processes.
Thus, in this paper we aim at investigating more in detail the link between large-amplitude electric field fluctuations and the emergence of bumps in the ion distribution function, as a a possible evidence of excitation of ion-bulk waves identified at small scales in the HVM simulations.

\section{Dataset and Analysis} \label{sec:method}
\begin{figure}
\centering
\includegraphics[width=12 cm]{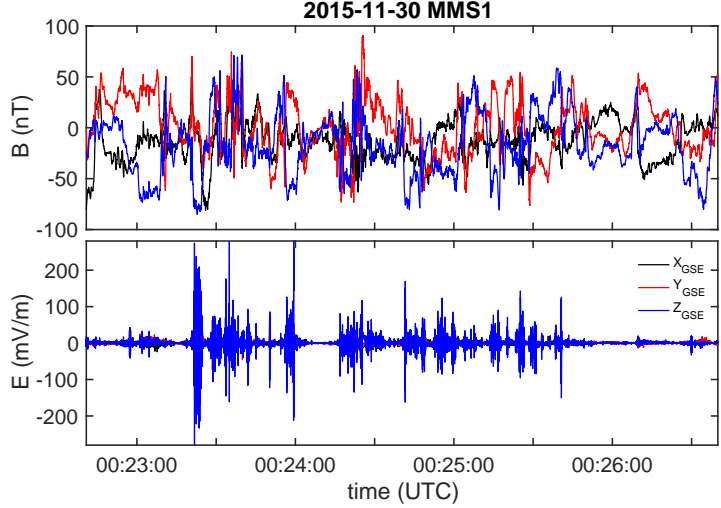}
\caption{Overview of the time series in GSE from MMS1 for the magnetic field vector components at $\sim 1$ kHz resolution (top panel) and for the electric field vector components at $8$ kHz (bottom panel). \label{fig:B-E}}
\end{figure}

We analyze one $5$-minute interval recorded from 00:21 to 00:26 UT on 2015 November 30, when the MMS spacecraft was in the quasi-parallel turbulent magnetosheath (see \citet{Yordanova16,Voroes17,Perri2020}). 
This interval is characterized by crossings of ion \citep{Voroes17} and electron~\citep{Eriksson16} diffusion regions of reconnection events, as well as multiple current sheets. A strong guide magnetic field $B_0\sim 44$ nT (averaged over the whole 5-minute interval) is present. The plasma $\beta$ is highly variable, and the ion-to-electron temperature ratio is $T_i/T_e \sim 7$. 
\citet{Perri2020} found in this interval strong correlation between the intensity of the electric field and the deviation from Maxwellian of the ion VDFs. 
In this work, we will explore the possible physical mechanism that gives rise to such correlation. 
To this aim, we make use of the high resolution (150 ms) ion VDFs from the Fast Plasma Investigation (FPI) instrument on board MMS \citep{Pollock16}, of the electric field data from Electric Double Probes (EDP) instrument (at about $8$ kHz sampling rate) \citep{Torbert16,Ergun16, Lindqvist16}, and of the magnetic field from the merged fluxgate (FGM) \citep{Russell16} and search coil (SCM) data \citep{LeContel16}, at about 1kHz resolution \citep{Fischer16}. 
An overview of the magnetic and electric field components in the geocentric-solar-ecliptic (GSE) reference frame is displayed in Figure \ref{fig:B-E}, where the highly fluctuating aspect of the magnetosheath region can be easily recognized.

\begin{figure*}
\centering
\includegraphics[width=6.9 cm]{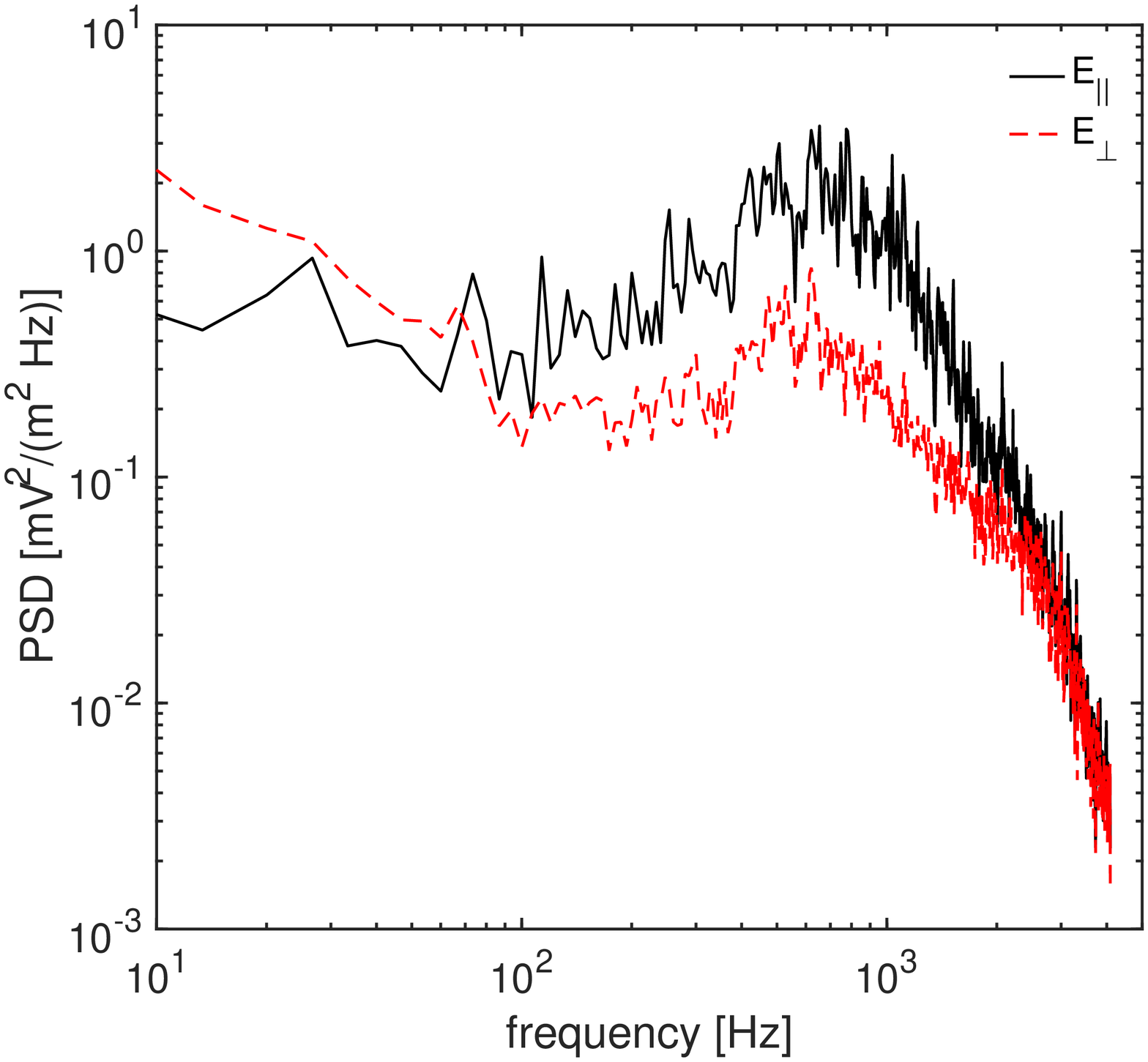}
\includegraphics[width=7.3 cm]{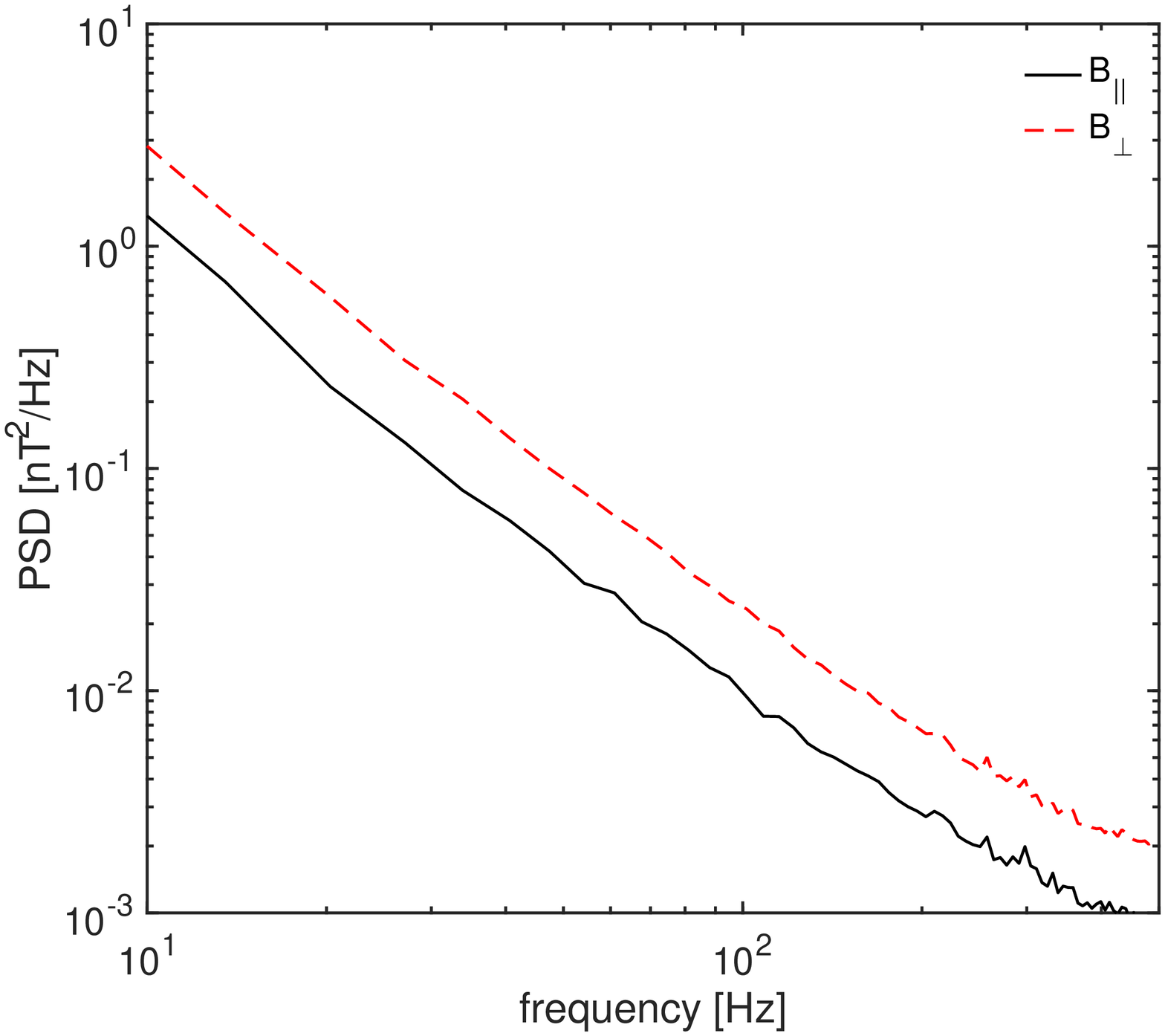}
\caption{Left panel: Spectrum of the parallel (solid black line) and perpendicular (dashed red line) components of the electric field fluctuations averaged over the selected electrostatic intervals (see text for details). There is a clear bump, higher for the parallel component, at about $500$ Hz. Right panel: Spectrum of the magnetic field fluctuations parallel (solid black line) and perpendicular (red dashed line) to the mean magnetic field direction, averaged over the same intervals. No magnetic features are detected. \label{fig:psd}}
\end{figure*}

\subsection{Evidence of high-frequency features in the electric field and ion-frequency signatures in the magnetic field}
Within the $5$-minute interval, we have computed the power spectral density (PSD) of the electric field fluctuations in the directions parallel and perpendicular to the mean magnetic field $\rm B_0$. The PSD has been computed using $150$ ms disjoint intervals. We have noticed that various intervals exhibit a clear peak or bump in the sub-proton range in the parallel electric power spectrum $\rm PSD(E_{||})$. This was also observed in numerical studies in regions where high parallel electrostatic activity at sub-proton scales has been found due to the growth of IBK waves \citep{Valentini11b}. We have then chosen such sub-intervals with a high-frequency feature (in the spacecraft frame) in $\rm PSD(E_{||})$. It is worth mentioning that many of such sub-intervals are characterized by $\nabla\times \mathbf{E} \sim 0$ ( as determined applying the four-spacecraft estimation of the $\nabla\times \mathbf{E}$ and choosing an arbitrary small threshold, so that $\nabla\times \mathbf{E}/\nabla \cdot \mathbf{E}< 0.5$), implying that they are indeed nearly electrostatic samples. We have actually found $128$ quasi-electrostatic sub-intervals, and computed the magnetic and electric field PSDs for each of them. 
Figure \ref{fig:psd} shows the PSD of the electric (left panel) and magnetic (right panel) field fluctuations, in the perpendicular (red dashed line) and parallel (solid black line) direction with respect to the local mean field. 
All spectra have been averaged over the $128$ selected sub-intervals. 
The parallel electric field PSD exhibits a bump at sub-ion frequencies ($\sim 500$ Hz), while there are no noticeable high-frequency features in the magnetic field PSD, implying the electrostatic nature of the fluctuations. 
Notice that the Nyquist frequency for the magnetic field data is very close to the (spacecraft frame) frequency of the $\rm PSD(E_{||})$ bump. However, $\rm PSD(B)$ keeps decreasing with the frequency, without any indication of deformation of the power-law scaling. 
Furthermore, on average the power in the $B_{\perp}$ fluctuations is higher than along the parallel direction, suggesting that energy is mostly accumulated along the transverse magnetic field fluctuations. Finally, a bump is also observed in the averaged $E_{\perp}$ spectra, although with lower amplitude than for the parallel component. This feature is persistently observed in all the selected intervals.

It is worth pointing out that \citet{Valentini14b} found similar PSD features using STEREO data, suggesting the presence of an electrostatic channel of fluctuations that transfers energy from large-scale Alfv\'enic fluctuations to smaller scales.
This energy transfer process has also been detected along the mean field in HVM simulations, at scales smaller than the proton inertial length \citep{Valentini08}, and has been attributed to the generation of a plateau (or bump) in the longitudinal VDF near the proton thermal speed (for plasma $\beta\sim 1$). This feature is caused by the resonant interaction of protons with left-handed polarized Alfv\'en cyclotron waves (ICW) at scales larger than the proton scales \citep{Valentini09}. 
The presence of this plateau excites the electrostatic IBK waves via an instability process that efficiently transfers energy from large to short scales in a non local way.

Thus, we have tried to find evidence, within portions of the signal characterized by high parallel electric field activity, of the IBK driver at the ion scales, namely left-handed polarized ICW. 
In order to do so, we have applied the wavelet transform coherence (WTC) analysis to identify coherent regions of the signal, where the phase relationship in time and time-scale (frequency) between two signals is fixed \citep[][]{Torrence1999,Grinsted2004,Lion2016}.
Considering the components of $\mathbf{B}$ perpendicular to the mean field, it is possible to define a cross-wavelet transform of $B_{\perp1}$ and $B_{\perp2}$ as $\mathcal{W}^{\perp1,\perp2}(f,t)=\mathcal{W}^{\perp1}(f,t)\mathcal{W}^{\perp2*}(f,t)$, being $\mathcal{W}$ the complex wavelet transform \citep[][]{Farge92} and $*$ the complex conjugation.
The coherence coefficient of the perpendicular components of the magnetic field is then computed as
\begin{equation}
    R^2(f,t)=\frac{|\mathcal{S}(f\mathcal{W}^{\perp1,\perp2}(f,t))|^2}{\mathcal{S}(f|\mathcal{W}^{\perp1}(f,t)|^2)\cdot \mathcal{S}(f|\mathcal{W}^{\perp2}(f,t)|^2)},
    \label{eqcoherence}
\end{equation}
where $\mathcal{S}(\mathcal{W}(f,t))=\mathcal{S}_f(\mathcal{S}_t(\mathcal{W}(f,t)))$ is a smoothing operator \citep[][]{Grinsted2004} in both time and frequency (the reader is deferred to \citet{Torrence1999,Grinsted2004} for the complete expressions of the smoothing operators in frequency and time, determined according to the Morlet wavelet transform). 
By definition, the coherence coefficient is confined between $0$ (no coherence) and $1$ (complete coherence). 
Additionally, we have tested the polarization of the high-coherence regions by computing the phase difference between the signals of the perpendicular components of the magnetic field. Indeed, $\phi(f,t)=\arg (\mathcal{W}(f,t))$ can be considered as the local phase of the signal and then $\Delta \Psi_{\perp1,\perp2}(f,t)=\phi_{\perp1}(f,t)-\phi_{\perp2}(f,t)$ \citep[][]{Grinsted2004,Lion2016} is the phase difference in the range $[-\pi,\pi]$. 
For positive local mean magnetic field, $\Delta \Psi_{\perp1,\perp2}(f,t)=0$ represents linear polarization, while $\Delta \Psi_{\perp1,\perp2}(f,t)= \pi/2 \; (-\pi/2)$ indicates right-hand (left-hand) circular polarization.

\begin{figure*}
\centering
\includegraphics[width=7 cm, height=7 cm]{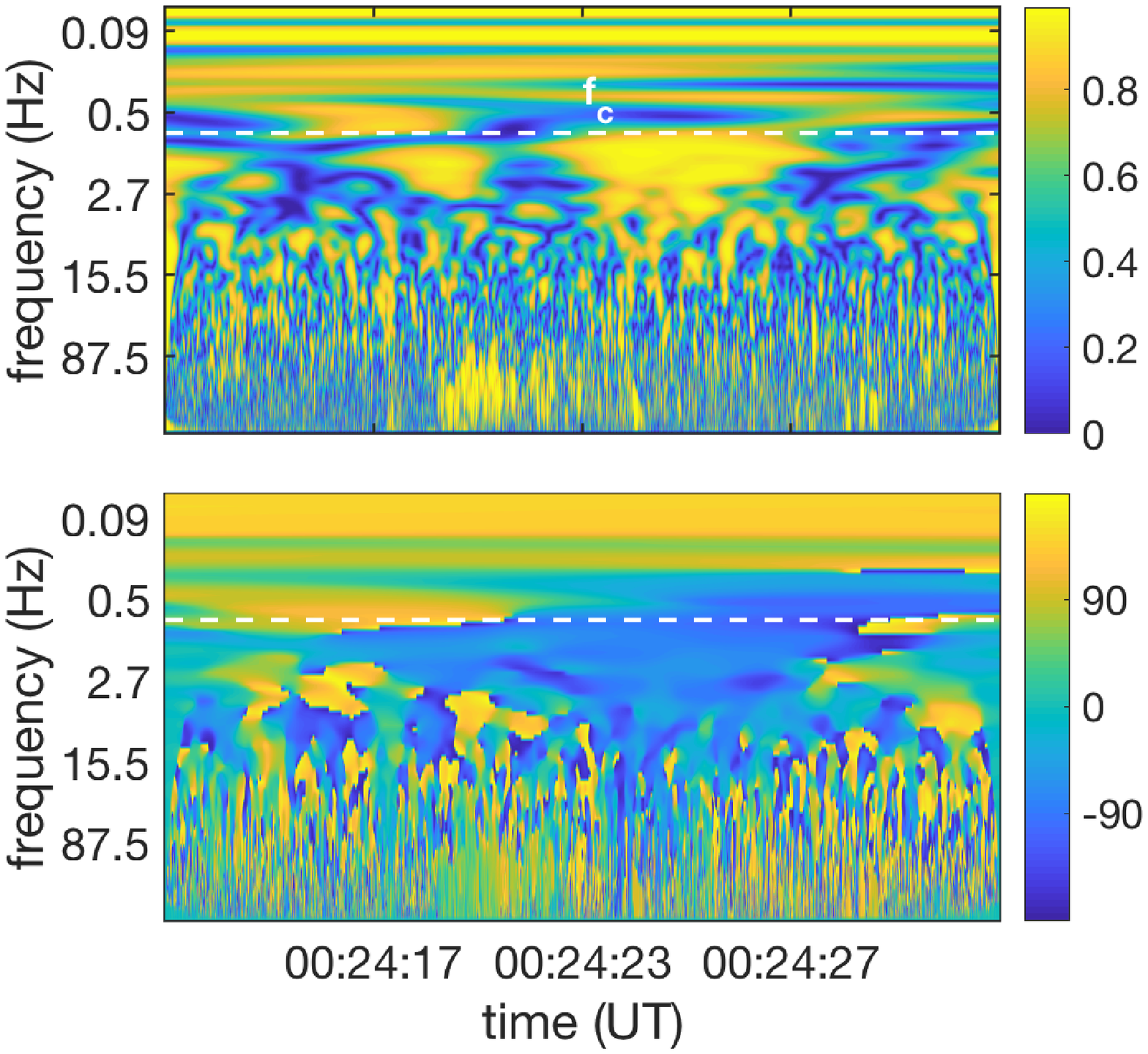}
\includegraphics[width=7 cm]{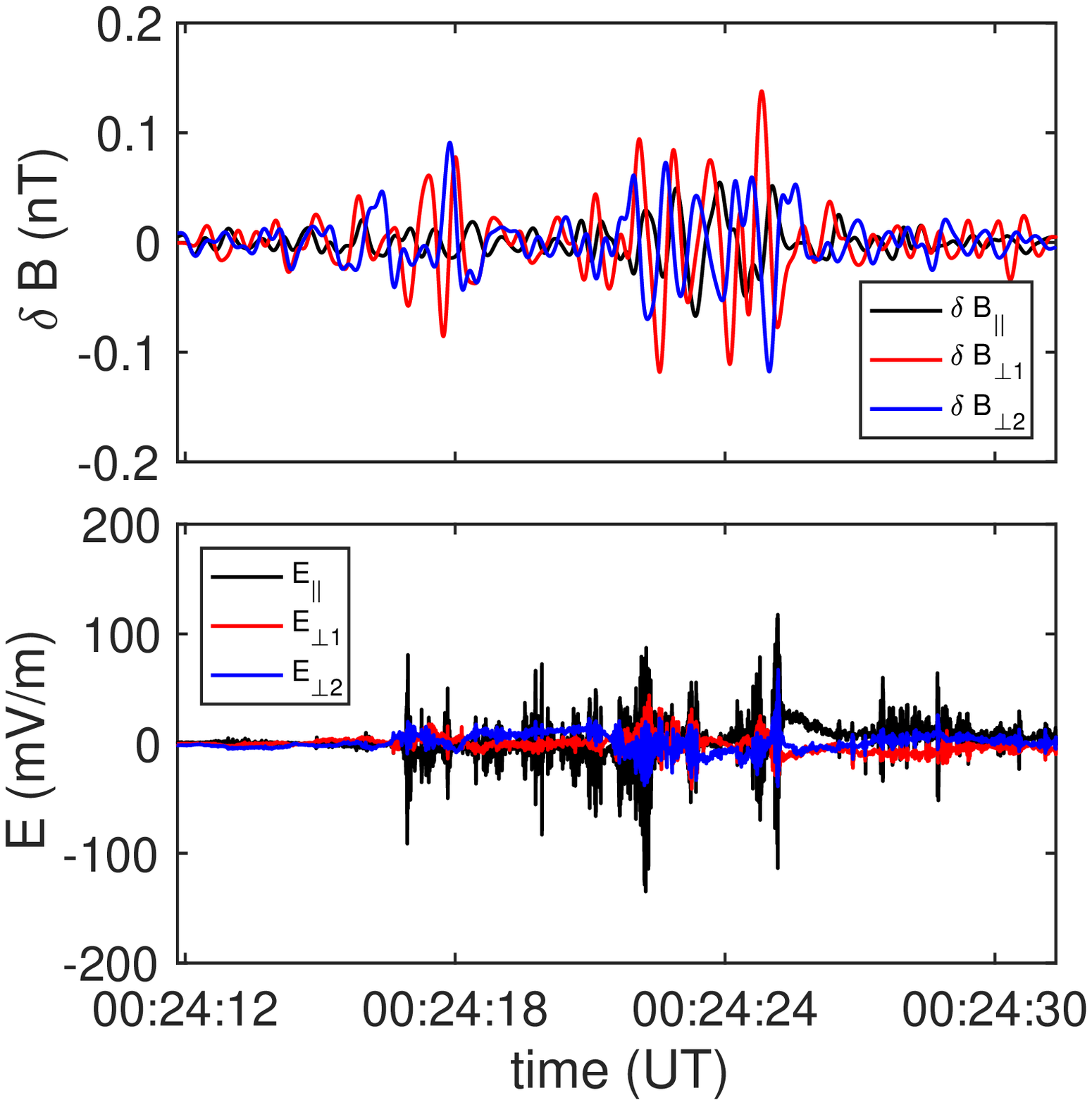}
\caption{Left: Spectrograms of the coherence of the perpendicular components of the magnetic field (top panel) and of their phase difference (bottom panel). The ion cyclotron frequency is indicated by the horizontal dashed line. See text for further details. Right: time series of the magnetic field fluctuations band-filtered between $0.6$ and $2.7$ Hz and rotated into the local mean field reference frame (top panel). Electric field time series rotated into the local mean field reference frame (bottom panel). \label{fig:ICW}}
\end{figure*}

Top-left panel of Figure \ref{fig:ICW} displays the scalograms of the coherence coefficient $R^2(f,t)$ of the components perpendicular to the local mean field for a $\sim 20$~s sub-interval with strong electric field high frequency fluctuations (in particular along the parallel component of $\mathbf{E}$, as can be seen in the bottom-right panel of Figure \ref{fig:ICW}). For a joint analysis between the magnetic field and the electric field time series, it has been necessary to analyze longer time intervals (which include several of the above mentioned $150$ ms electrostatic intervals) in order to detected ion scale wave activity in the magnetic field. Indeed, the coherence analysis unveils the presence of highly coherent regions in the middle of the sub-interval, near the proton gyrofrequency $f_c$, indicated by the horizontal dashed-white line in the scalogram. 
Such high coherence regions also display a phase difference close to $-\pi/2$ (bottom-left panel of Figure \ref{fig:ICW}). Hence, ion-scale left-handed polarized  coherent events in the magnetic field are observed in correspondence with rapid sub-proton large-amplitude fluctuations of the parallel electric field.

To highlight the nature of the magnetic fluctuations around the ion cyclotron frequency, we have band-pass filtered the magnetic field components parallel and perpendicular to the local $\mathbf{B}_0$ within the selected interval, by means of wavelet transform $\mathcal{W}_i (\tau_j,t)$ \citep{Torrence1998,Perrone2016,Perrone2017,Perrone2020}, namely
\begin{eqnarray}
\label{eq:fluct}
    \delta b_i(t) = \frac{\delta j \delta t^{1/2}}{C_\delta \psi_0(0)} \sum_{j=j_1}^{j_2} \frac{\mathcal{R} \left[ \mathcal{W}_i (\tau_j,t)\right]}{\tau_j^{1/2}}
\end{eqnarray}
where $\mathcal{R}$ represents the real part, $j$ is the scale index, $\delta j$ is the fixed scale step, and $\tau_j$ is the time scale. We have used a Morlet wavelet~\citep{Torrence1998}, so that $\psi_0(0)=\pi^{1/4}$ and $C_\delta=0.776$. The limits of the band-pass filter adopted here are $\tau(j_1)=0.37$~s and $\tau(j_2)=1.66$~s, with $\tau =1/f$. 
The filtered magnetic fluctuations are shown in the top-right panel of Figure \ref{fig:ICW}, where indeed wave packets can be recognized. 
Notice also that the transverse fluctuations tend to dominate with respect to the parallel ones (black line), although the parallel component is not totally negligible; on the other hand, the magnetosheath interval analyzed is highly turbulent and fluctuating, so that compressive effects cannot be totally ruled out. In correspondence of such magnetic field wave packets, intense parallel electric field activity is observed at very high frequency, resulting in the features in the $\rm PSD(E_{||})$ around $500$ Hz. We have additionally found a high level of coherence at the ion timescale between the perpendicular 
fluctuations of both magnetic and electric field, which tend to be aligned or anti-aligned (not shown), in good agreement with a ICW event detected by \citet{He19} also in the magnetosheath.

\begin{figure*}
\centering
\includegraphics[width=7 cm]{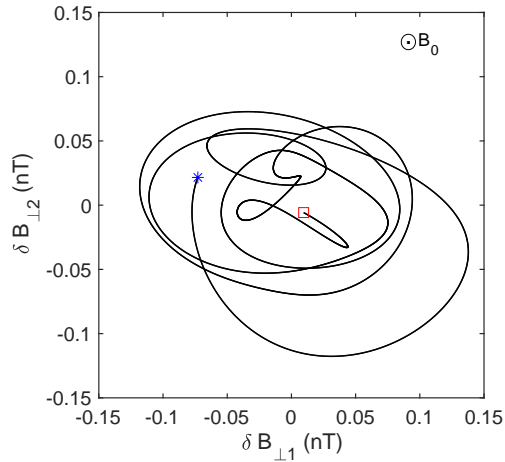}
\caption{Hodogram of the band-passed perpendicular fluctuations during the interval shown in Figure \ref{fig:ICW}. The red square indicates the starting point, while the blue star the ending point. Since the mean field $B_0$ is positive the clock-wise rotation indicates left-handed polarized waves.\label{fig:odogram}}
\end{figure*}

Once the possible ``driving" waves have been detected, we have plotted the hodogram of the perpendicular components of $\mathbf{B}$ in Figure \ref{fig:odogram}, within the main wave packet in the band-pass filtered magnetic field (at about 00:24:24 UT). The chosen packet corresponds to left-handed circularly polarized waves (the local mean field is positively directed and the versus of the hodogram is clock-wise, from the red square towards the blue star). Thus, the emerging picture is that high-amplitude bursts at high frequency in the parallel electric field can be in some intervals associated to the presence of left-handed polarized ICWs.

\subsection{ICWs from the dispersion relation using magnetic field time series}

In order to better validate the presence of ICWs at ion timescales during intervals of high-frequency parallel electrostatic activity, we have used magnetic field data to estimate a dispersion relation. Indeed, although the analyzed magnetosheath medium is highly turbulent and characterized by non-linear phenomena, we would like to check whether the presence of waves can leave a "linear memory" in the magnetic field fluctuations.  
Boosted by the launch of multi-satellite missions, several methods have been developed for determining magnitude and direction of the wavevector $\mathbf{k}$. These include minimum variance analysis \citep[][]{Sonnerup98}, multi-spacecraft phase differences \citep{DudokdeWit1995,Balikhin2003,Walker2004}, and k-filtering \citep{Pincon91,Sahraoui06}.
In this work, we adopt the single-satellite technique described in \citet{Bellan16}, which uses the Ampere's law $\mu_0 \mathbf{J}(\omega)= i\mathbf{k}(\omega)\times \mathbf{B}(\omega)$ under the hypothesis of time-dependent magnetic field and current density given by a sum of plane waves at a given position $\mathbf{x}$. 
The main limitation of this technique is that it associates a unique $\mathbf{k}$ to each frequency $\omega$. Thus, after straightforward calculations, it is possible to obtain the wave vector as
\begin{equation}
    \mathbf{k}(\omega)=i \mu_0 \frac{\mathbf{J}(\omega)\times \mathbf{B}^*(\omega)}{\mathbf{B}(\omega)\cdot \mathbf{B}^*(\omega)} \, .
    \label{eqkappa}
\end{equation}

\begin{figure}
\centering
\includegraphics[width=9 cm]{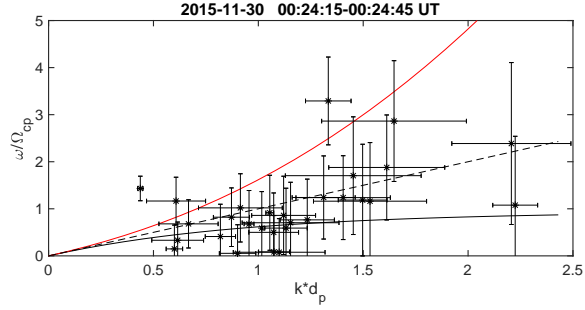}
\caption{The dispersion relation obtained from the magnetic field data within the MMS interval displayed in Figure \ref{fig:ICW} using a phase difference technique (see text). Error bars are also displayed. The dispersion relation curves for the ICWs (black solid line), for the Whistler branch (red solid line), and for the Alfv\'en waves (dashed line) are shown. \label{fig:dispersion}}
\end{figure}

The relative error on $\mathbf{k}(\omega)$, calculated here from the MMS1 measurements and reported in Figure \ref{fig:dispersion} with error bars, is determined as 
${\rm Err}=|(|\mathbf{J}^*|-|\mathbf{J}|)|/(|\mathbf{J}^*|+|\mathbf{J}|)$ \citep[][]{Bellan16}, where $\mu_0\mathbf{J}^*=i\mathbf{k}\times \mathbf{B}$ is the current density estimated using the k-vector from Equation \ref{eqkappa}. $\mathbf{J}(\omega)$ is computed from the FPI particle measurements, where the ion velocity is resampled onto the electron time tags, namely $\mathbf{J}=n_e\; q(\mathbf{V}_i-\mathbf{V}_e)$. The mean and the standard deviation of the angle between the k vector and the magnetic field direction is $\theta_{kb}\sim 45\pm 15^{\circ}$.

We have performed the k-vector determination for the sub-interval shown in Figure \ref{fig:ICW}, where large amplitude/high-frequency electrostatic waves were clearly identified. 
Note that in this sub-interval the ion bulk speed exhibits rapid fluctuations. This gives rise to large errors in the Doppler-shift determination of the frequencies in the plasma frame, being $\omega_{\rm s/c}=\omega+(\mathbf{k}\cdot \mathbf{V}_{\rm bulk})/{\rm 2\pi}$.
The errors on $\omega$ have been estimated by using the above k-vector and finding the component of the velocity in the direction of $\mathbf{k}$. The standard deviation of the velocity is calculated and multiplied by k (in angular units) to give the errors.
The obtained $k-\omega$ relation, with the relevant error bars described above, is shown in Figure \ref{fig:dispersion}, where the wavenumbers have been normalized to the proton inertial length and the frequencies to the proton cyclotron frequency. 
Along with the data points, we have plotted in Figure \ref{fig:dispersion} the dispersion relation curves for the ICWs (black solid line), for the Whistler waves (red solid line), and for the Alfv\'en waves (dashed line). The dispersion relation at low $k-{\rm s}$ gives indication of the presence of ICWs, which can be the driver of the IBK waves at larger wave numbers, leaving a trace of the linear counterpart in such a turbulent plasma sample. 

\section{Features in the observed ion VDFs}
We now focus on the possible wave-particle interaction processes generating the bumps in the electric field parallel spectrum. To this aim, we have analyzed the ion VDFs for intervals with and without the electric spectral bumps around $500$ Hz. 
In the absence of the bump, we observed either broadening of the ion VDFs, corresponding mainly to particle heating, or anisotropic core distribution in the plane ($\rm \mathbf{B}$, $\rm \mathbf{E}\times \mathbf{B}$) \citep{Voroes17}. Such anisotropic feature consists of a typical D-shape distribution with a cut-off in the anti-field-aligned direction (see Figure 13 in \citet{Voroes17}), and is typical of reconnection outflow regions. 
In some occasions, shift of the core VDF population was also observed in the $\mathbf{E} \times \mathbf{B}$ direction, corresponding to the convection direction of ions. 

\begin{figure}
\centering
\includegraphics[width=9 cm]{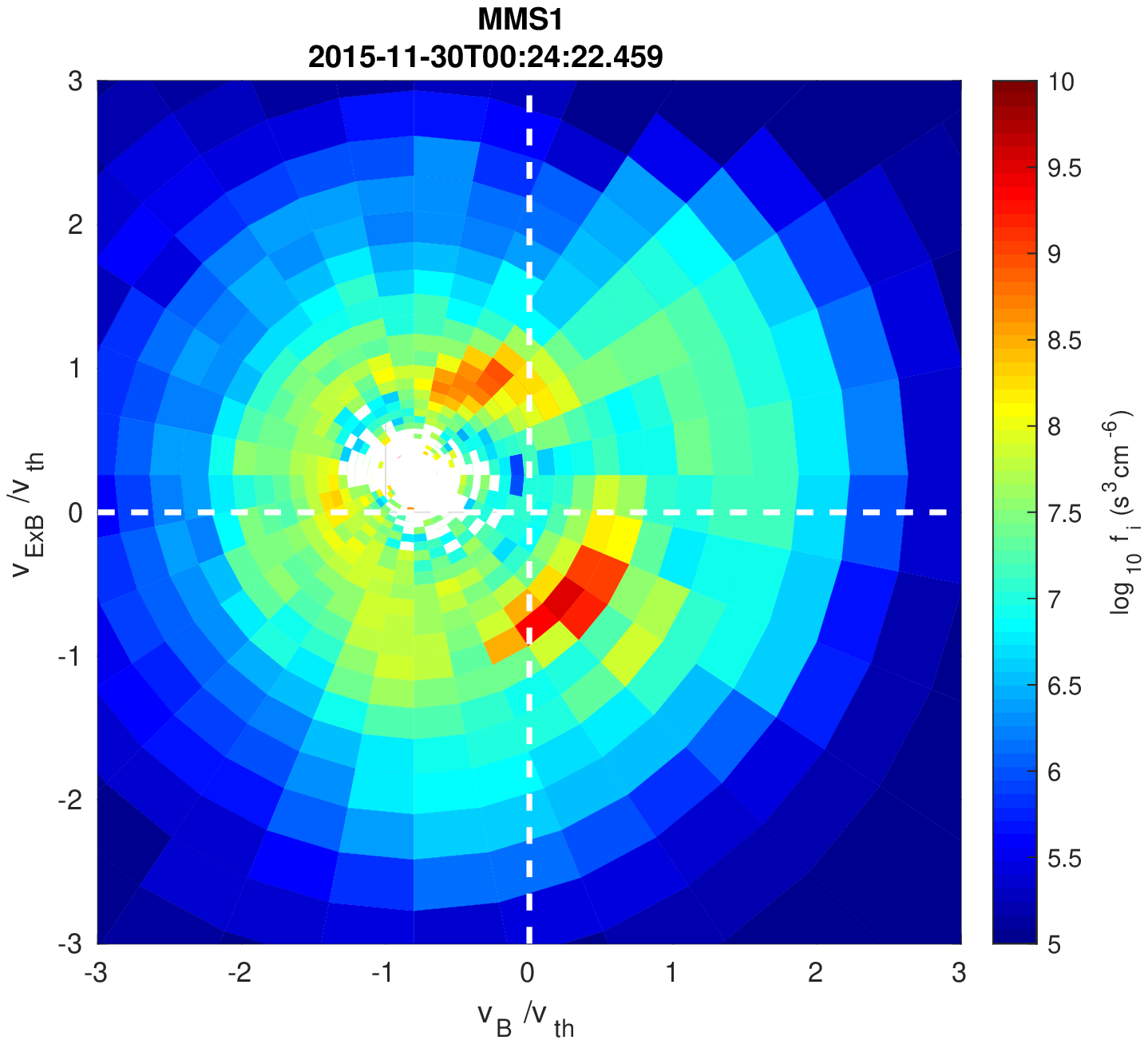}
\includegraphics[width=9 cm]{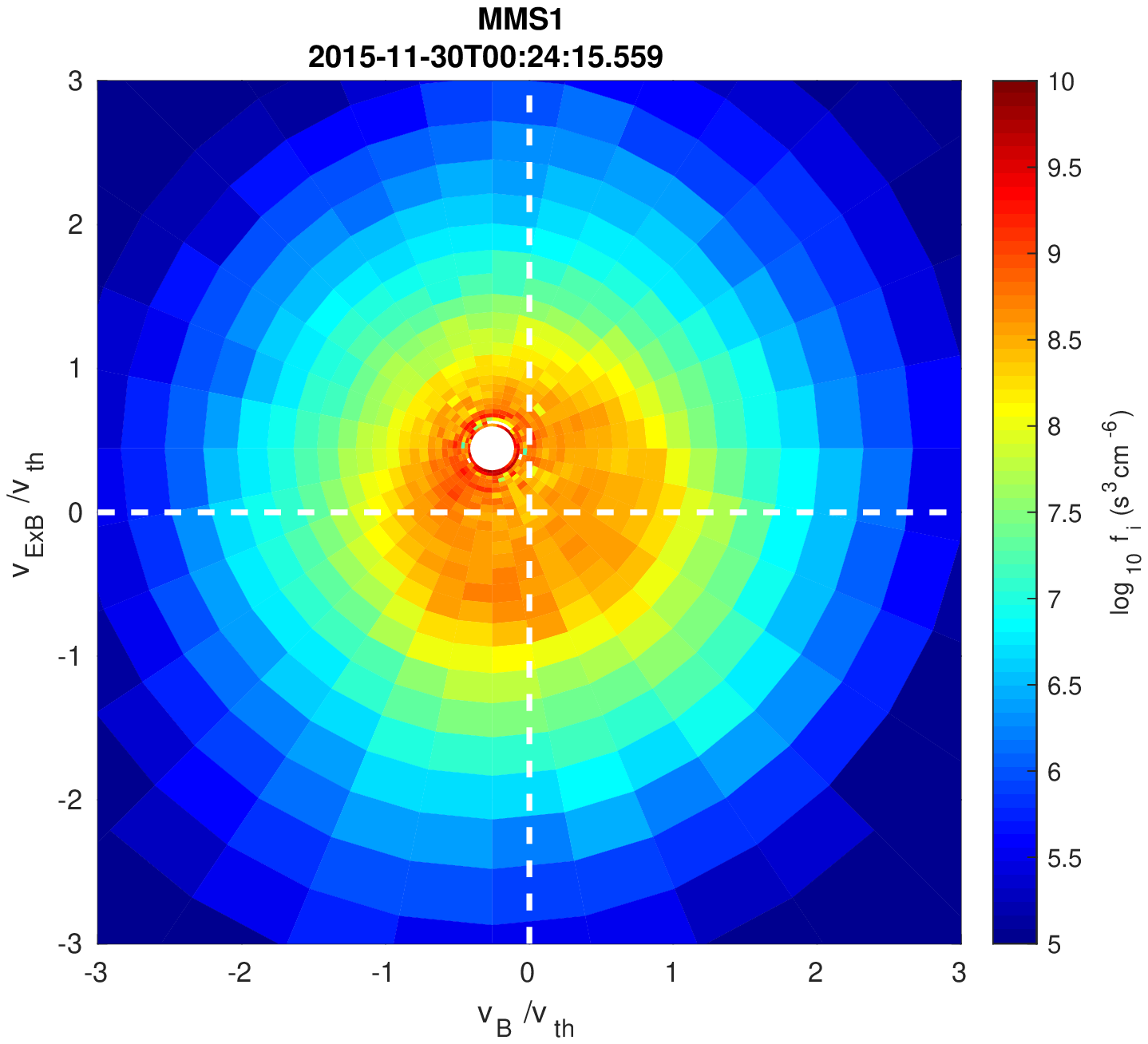}
\caption{Two-dimensional VDF cut (in the plasma frame) in the plane ($\rm \mathbf{B}$, $\rm \mathbf{E}\times \mathbf{B}$) measured almost in the middle of the interval in Figure \ref{fig:ICW} (top panel), and at the beginning of the same interval where locally no bump in the $\rm PSD(E_{||})$ has been observed (bottom panel).  Velocity values are normalized to the local ion thermal speed. \label{fig:vdf}}
\end{figure}

On the other hand, in presence of the parallel electrostatic activity described in this work, the ion VDFs exhibit different characteristics. 
Figure \ref{fig:vdf} shows the 2D cuts of the MMS1 ion VDFs, projected onto the ($\rm \mathbf{B}$, $\rm \mathbf{E}\times \mathbf{B}$) plane in the plasma frame, in the middle of the interval reported in Figure \ref{fig:ICW} (top panel) and at the beginning of the same interval where very low electrostatic activity along the parallel direction is observed with no bump in the $\rm PSD(E_{||})$ (bottom panel). The $\rm \mathbf{E}\times \mathbf{B}$ direction has been estimated using the average fields directions within time windows of $0.15$ sec centered on the times indicated in Figure \ref{fig:vdf}. The two cuts show very different shapes: an ion beam can be easily identified in the ion VDF relevant to the Figure \ref{fig:ICW} sub-interval travelling at a speed comparable with the ion thermal speed $v_{th,i}$ along the magnetic field direction. Similar features have been identified in the Earth's magnetosphere in \citet{Sorriso2019}. Conversely, in the 2D cut displayed in the bottom panel of Figure \ref{fig:vdf} no particular features can be identified and the VDF tends to be almost isotropic within the ($\rm \mathbf{B}$, $\rm \mathbf{E}\times \mathbf{B}$) plane. A visual inspection of the ion VDFs in this $\sim 20$ s sub-interval has highlighted shapes similar to that in the top panel of Figure \ref{fig:vdf} within regions with $E_{||}$ much larger than $E_{\perp1}$ and $E_{\perp1}$ (condition that is basically found when ICW packets are detected at $f\sim f_c$); ion VDFs with almost isotropic shape at the beginning of this sub-interval when both $\mathbf{E}$ and $\delta\mathbf{B}$ fluctuations are low; VDFs dominated by a shifted core along the $\rm \mathbf{E}\times \mathbf{B}$ direction at times $\rm t>00:24:25 UT$ when the amplitude of the $E_{\perp}$ components starts becoming comparable with that of $E_{||}$.
Thus, the formation of beams (and then plateau) in the ion VDFs can be related to the presence of an enhancement in the parallel electric field fluctuation energy at sub-ion scales, possibly driven by the ICWs at ion scales. 
All these ingredients suggest the development of electrostatic IBK waves via an instability  caused by the resonant interaction of ions with left-handed polarized ICWs. 
Indeed, the presence of a region with positive derivative in the parallel velocity of the ion VDFs causes a beam-plasma type instability. Such an instability occurs in regions where the ions are trapped in the wave potential well, leading to a non local transport of the electrostatic fluctuation energy that gives rise to a bump in the energy spectrum of the parallel electric field fluctuations \citep{Valentini11c}.
As stated above, such an interaction has been found in HVM simulations \citep{Valentini08}, generating a bump in the ion VDFs in the longitudinal direction at about $v_{th,i}$ for plasmas with $\beta\sim 1$. In the analyzed sub-interval in Figure \ref{fig:ICW} the proton plasma beta tends to be $\beta<5$, so that it is close to the value used in the numerical simulations.

\begin{figure}
\centering
\includegraphics[width=9 cm]{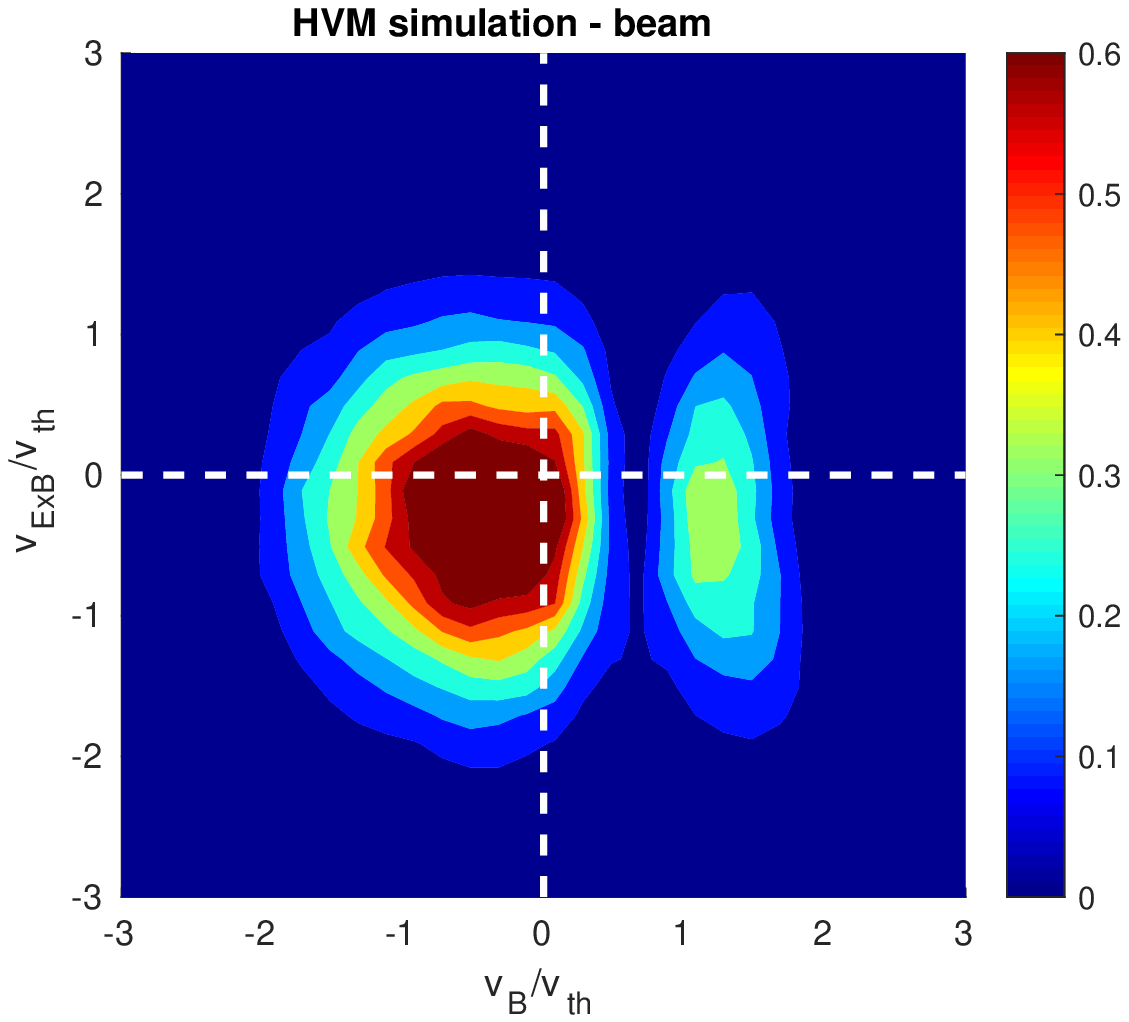}
\includegraphics[width=9 cm]{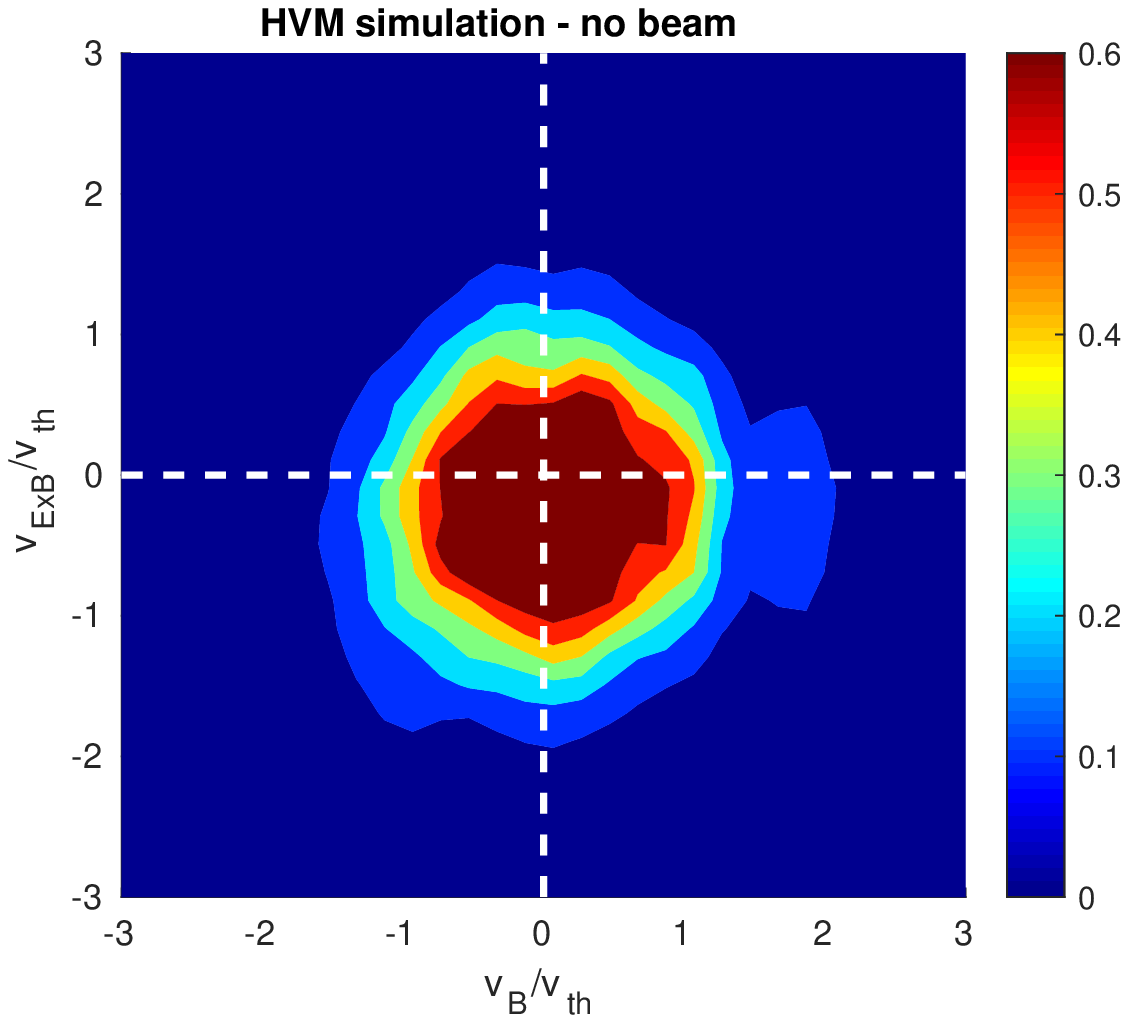}
\caption{Two-dimensional VDF cut in the plane ($\rm \mathbf{B}$, $\rm \mathbf{E}\times \mathbf{B}$) within a region in the simulation box where a beam in the proton VDF emerges and large amplitude spikes in the parallel electric field are detected (top panel) \citep[][]{Valentini11}, and within a region where no electrostatic activity associate to IBK waves has been excited (bottom panel).  Velocity values are normalized to the simulation ion thermal speed. \label{fig:vdfsim}}
\end{figure}

\subsection{Comparison with HVM simulations}
Kinetic simulations have shown that IBK electrostatic fluctuations can be present at scales smaller than the proton inertial length, and propagate roughly at the proton thermal speed \citep{Valentini08,Valentini09,Valentini11b}. 
Such novel branch of waves has been investigated thoroughly for different temperature ratios in the range $T_e/T_p=[1,10]$. It was shown that, unlike the ion acoustic waves that are strongly affected by Landau damping~\citep[as predicted by the linear theory,][]{KrallTrivelpiece1986}, the IBK waves are not completely damped. 
We have run 1D HVM simulation~\citep{Valentini07} (not shown) with realistic electron-to-proton mass ratio in order to check whether IBK waves can be excited also in the presence of cold electrons. The simulation indeed aims at sharing plasma properties similar to those observed within the MMS magnetosheath interval here analyzed, where a very low electron-to-proton temperature ratio is observed, i.e., $T_e/T_p \sim 0.14$. 
In the HVM simulations, clear signatures of electrostatic fluctuations have been observed for initial temperature ratio $T_e/T_p = 0.8$ and proton plasma $\beta_p = 2v_{th,p}^2/v_A^2 = 0.5$. 
For simulation times above $t\,\omega_{cp} \simeq 50$, the electric energy spectra show the presence of electrostatics waves in the tail. 
A train of resonant modes starts to grow in the wavevector range $k d_p = [8, 30] $, associated with spikes in the parallel electric field. Such large amplitude fluctuations of the parallel electric field are similar to those observed in some portions of the magnetosheath interval (see the bottom-right panel of Figure \ref{fig:ICW}). 
Spikes in $E_{||}$ are associated with the presence of a small beam in the longitudinal proton VDF, which originates from the interaction between protons and left-handed ICWs. 
An example of the proton VDF in the proximity of the region presenting electric field spikes in the numerical simulation is reported in Figure \ref{fig:vdfsim} (top panel).
Notice the similarity with the observed VDF in the top panel of Figure \ref{fig:vdf}, although nonlinear effects occurring in the magnetosheath environment tend to further distort the VDF \citep{Greco12,Valentini16}, leading to a more diffusive particle velocity distribution in real data. A proton VDF within a region in the simulation box (taken at the same running time) where no features in the electric field are observed is also displayed for comparison in the bottom panel of Figure \ref{fig:vdfsim}.

At later times in the simulation, the IBK features in the electric field and the bump in the high-frequency spectrum are no longer visible. This suggests that Landau damping eventually comes into play effectively for this low value of $T_e/T_p$.
For smaller values of the temperature ratio, electrostatic activity is still observed, but IBK waves are quickly damped. 

Therefore, it is possible to expect the presence of IBK waves also in an environment with properties similar to the terrestrial magnetosheath.

\section{Conclusions}\label{sec:conclusions}
We have investigated the kinetic processes that give rise to an enhancement of the parallel electric field fluctuations at very high frequencies in the spacecraft frame within the terrestrial magnetosheath. Spikes in the parallel component of the electric field has been observed to emerge at high wave numbers in HVM numerical simulations \citep[][]{Valentini11b}, owing to the presence of a diffusive plateau in the longitudinal proton VDF at about the proton thermal speed (for $\beta\sim 1$). Such a plateau forms because of the interaction of thermal protons with left-handed polarized ICWs. Once
the plateau has been created, a novel electrostatic branch of longitudinal IBK waves is developed, even for low values of the electron to ion temperature ratio. This branch gives rise to a new channels for the development of turbulence towards small scales \citep[][]{Valentini11}. Thus, using high resolution magnetic field and electric field measurements from MMS1, we have selected $0.15$ s intervals where the spectrum of the electric field fluctuations parallel to the mean magnetic field exhibits a clear bump at about $500$ Hz and no features in the magnetic field power spectrum, both along and transverse to the mean field, have been detected within the kinetic (sub-proton) range. An analysis of coherence and of phase difference applied to the perpendicular magnetic field components, using wavelet transforms, has allowed us to localize the driver of such an intense electrostatic activity at smaller (ion) frequencies. In several sub-intervals selected, a clear signature of the presence of ICWs at ion scales has been observed since:
\begin{itemize}
    \item the coherent coefficient of the perpendicular components is very high and well localized around the ion cyclotron frequency and within the time intervals of large amplitude fluctuations in the parallel electric field;
    \item the phase difference between the perpendicular components tends to be close to $90^\circ$ within high coherence regions;
    \item on average the power stored within the magnetic field perpendicular components is higher than that in the parallel component;
    \item a band-pass filtering of the magnetic field fluctuations around the ion cyclotron frequency shows well defined wave packets that emerge in correspondence of the high-frequency wave packets in the electric field signal;
    \item band-pass filtered magnetic fluctuations have larger amplitude in the perpendicular components and tend to be left-handed circularly polarized;
    \item the single spacecraft technique developed in \citet{Bellan16} has allowed us to determine a dispersion relation from the magnetic field time series that, within the electrostatic interval chosen, is reminiscent of the dispersion relation of ICWs, despite the high variability of the parameters that characterize such a magnetosheath data set. 
\end{itemize}
Further, the analysis of the ion VDFs shows the development of a beam along the local mean field direction and located at about the local ion thermal speed in the plasma frame. This is in very good agreement with the VDFs from HVM numerical simulations with low electron-to-proton temperature ratio within regions of large-amplitude spikes in the parallel electric field. 
All the above evidences claim for the development of a new channel of turbulence transfer from large to small scales within the kinetic range, already observed in numerical simulations \citep[][]{Valentini11, Valentini11b} and previously observed using STEREO observations in the solar wind \citep{Valentini14b}.
In summary, we propose the following sequence of steps that give rise to such a cascade channel:
\begin{itemize}
    \item The magnetohydrodynamic turbulent energy cascade populates the branch of ion-cyclotron fluctuations at frequency of the order of the ion-cyclotron frequency;
    \item Resonant interaction of ion-cyclotron fluctuations (propagating parallel to the background field) with ions leads to the generation of a diffusive velocity plateau \citep[][]{KennelEngelmann1966} in the longitudinal ion distribution function; this plateau, for ion beta of order unity, is located in the core of the distribution;
    \item Once this plateau has been created, the longitudinal ion-bulk wave channel can be excited \citep{Valentini09,Valentini11} and is available for turbulence to develop towards smaller wavelengths (higher frequencies), along an acoustic-like dispersion relation \citep{Valentini11b}.
\end{itemize}

We would like to remark that we were able to isolate evidence of the high frequency IBK waves and of the low frequency left-handed polarized ICWs in a very complex and highly structured environment as the Earth's magnetosheath. 
This has been possible thanks to both the high resolution of magnetic field, electric field, and plasma measurements from MMS, and also to the full angular coverage of the particle distribution function from FPI.
The solar-wind plasma would be actually ideal for such a study and the plasma properties would be also closer to the numerical experiments in \citet{Valentini11,Valentini11b}. It would also be relevant to evaluate the dissipation of magnetic energy \citep{He19} through the branch that transfers turbulent energy from ICWs towards IBK waves. We plan to investigate the presence of such a branch of magnetic energy transfer from large to small scales in the inner heliosphere by using the high cadence instruments on board the Solar Orbiter mission.

\begin{acknowledgments}
This work has received funding from the European Unions Horizon 2020 research and innovation programme under grant agreement No 776262 (AIDA, www.aida-space.eu). EY and LSV were supported by the Swedish Civil Contingencies Agency, grant 2016-2102. LSV was supported by SNSA grant 86/20. 
\end{acknowledgments}


\bibliography{biblio}{}
\bibliographystyle{aasjournal}


\end{document}